# Astrophysics with Radioactive Atomic Nuclei

**Executive Summary.** We propose to advance investigations of electromagnetic radiation originating in atomic nuclei beyond its current infancy to a true astronomy. Such nuclear emission is independent from conditions of gas, thus complements more traditional astronomical methods used to probe the nearby universe. Radioactive gamma-rays arise from isotopes which are made in specific locations inside massive stars, their decay in interstellar space traces an otherwise not directly observable hot and tenuous phase of the ISM, which is crucial for feedback from massive stars. Its intrinsic 'clocks' can measure characteristic times of processes within the ISM. Frontier questions that can be addressed with studies in this field are the complex interiors of massive stars and supernovae which are key agents in galactic dynamics and chemical evolution, the history of star-forming and supernova activity affecting our solar-system environment, and explorations of occulted and inaccessible regions of young stellar nurseries in our Galaxy. This White paper addresses Science Areas "Stars and Stellar Evolution (SSE)" and "The Galactic Neighbourhood (GAN)" of the US National Academy's Decadal Survey Outline Structure.

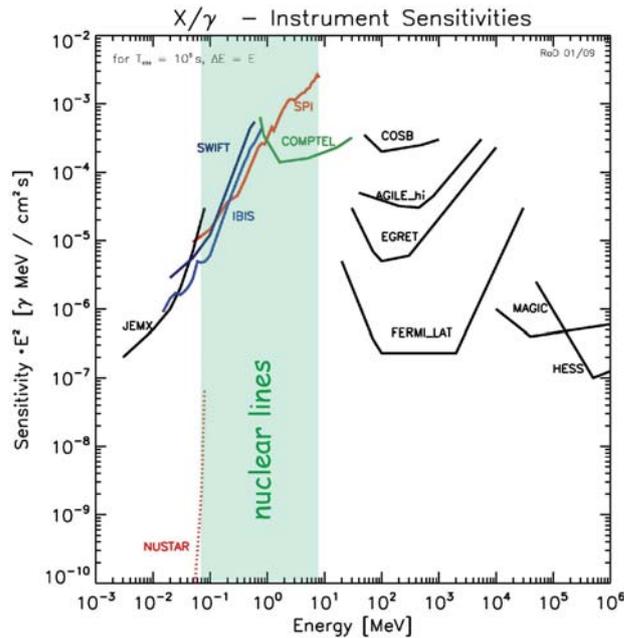

Fig. 1: *Instrumental sensitivities at supra-thermal energies*

| Isotope | Mean Decay Time | Decay Chain | γ-Ray Energy [keV] | Detected Source | Source Type |
|---|---|---|---|---|---|
| $^7$Be | 77 d | $^7$Be → $^7$Li* | 478 | (none) | Novae |
| $^{56}$Ni | 111 d | $^{56}$Ni → $^{56}$Co* → $^{56}$Fe*+e$^+$ | 158, 812; 847, 1238 | SN1987A, SN1991T(?) | Supernovae |
| $^{57}$Ni | 390 d | $^{57}$Co → $^{57}$Fe* | 122 | SN1987A | Supernovae |
| $^{22}$Na | 3.8 y | $^{22}$Na → $^{22}$Ne* + e$^+$ | 1275 | (none) | Novae |
| $^{44}$Ti | 85 y | $^{44}$Ti → $^{44}$Sc* → $^{44}$Ca*+e$^+$ | 78, 68; 1157 | SNR Cas A | Supernovae |
| $^{26}$Al | 1.04 10$^6$ y | $^{26}$Al → $^{26}$Mg* + e$^+$ | 1809 | Galactic Plane Regions | Stars, Novae Supernovae |
| $^{60}$Fe | 3.5 10$^6$ y | $^{60}$Fe → $^{60}$Co* → $^{60}$Ni* | 59, 1173, 1332 | Galaxy | Supernovae, Stars |
| e$^+$ | 10$^5$...10$^7$ y | e$^+$+e$^-$ → Ps → γγ.. | 511, <511 | Galactic Bulge, Disk(?) | Supernovae, Novae, … |

Table 1: *Spectral lines of relevance to astrophysics with radioactive nuclei*

*Contributing Authors:*  R. Diehl[1], P. von Ballmoos[2], S. Boggs[3], A. Burkert[4], A. Chieffi[5], N. Gehrels[6], J. Greiner[1], D. H. Hartmann[7], G. Kanbach[1], G. Meynet[8], N. Prantzos[9], J. Ryan[10], F.K. Thielemann[11], H. Zinnecker[12]

*with Affiliations:*  [1]*MPE Garching,* [2]*CESR/UPS Toulouse,* [3]*UC Berkeley,* [4]*LMU Munich,* [5]*IASF Rome,* [6]*GSFC Greenbelt,* [7]*Clemson University,* [8]*Geneva Observatory,* [9]*IAP Paris,* [10]*Univ. of New Hampshire,* [11]*Uni Basel,* [12]*AIP Potsdam*

**Astrophysical Issue:       The Interiors of Massive Stars and Supernovae**

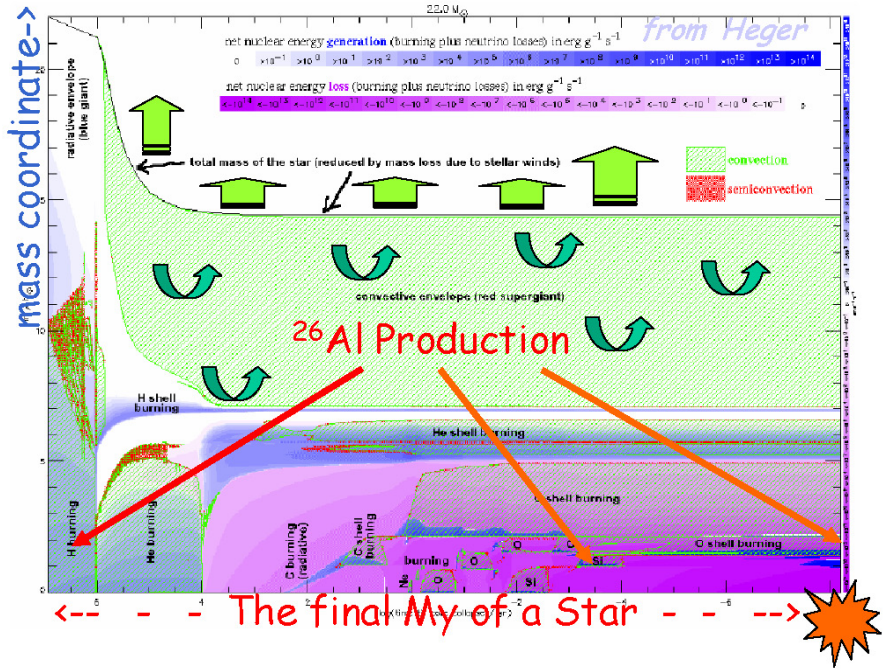

Fig. 2: *Structure of a massive star, as it evolves up to the terminal supernova (log. time scale) (adapted from Heger)*

Although we understand the basic processes that control the lives of massive stars, quantitative and robust models of their structure and evolution are still missing, which are key to properly treat and understand their roles in chemical evolution and in the important terminating events, the supernovae and gamma-ray bursts. For example, energy generation and transport within a massive star are results of subtle effects of material transport and mixing – driven by a variety of multi-scale agents such as buoyancy, differential rotation, radial oscillations, convective-flow moments of inertia – and are still mimicked by quite crude *approximations*, such as *semi-convection* and *mixing-length theory*. The complex sequential stages of shell-burning which characterize the late evolution of massive stars after core Helium burning (Fig. 2) are still very uncertain. Beyond, even the $^{12}C(\alpha,\gamma)$ reaction rate is too uncertain to allow a reliable computation of the advanced burnings and of the final Mass-Radius relation, which determines the *initial conditions* for core-collapse, the characteristics of the supernova explosion itself, and the subsequent propagation of the shock through the stellar envelope. As a result, nucleosynthetic yields from massive stars are uncertain by large factors even for a given stellar mass, and display also significant variations with stellar mass (Fig. 3).

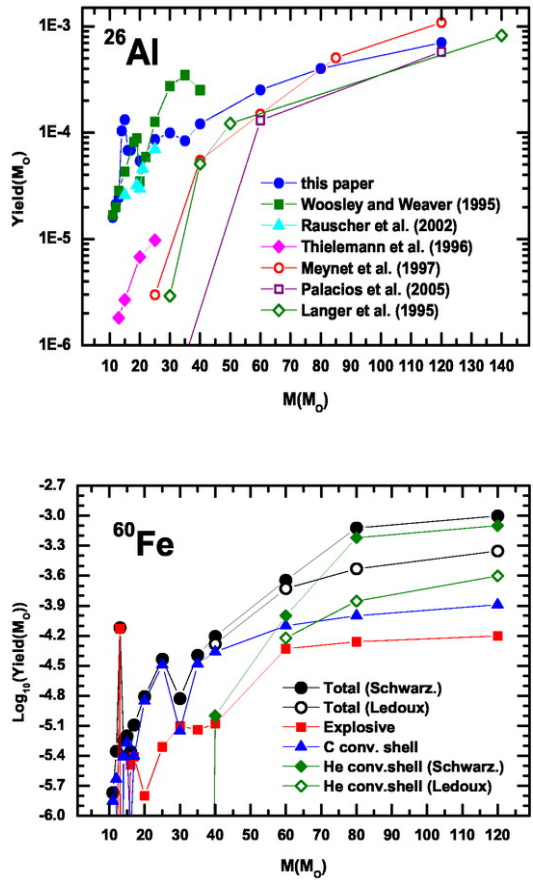

Fig. 3: *Model Yields versus stellar mass for $^{26}Al$ (top) and $^{60}Fe$ (bottom), illustrating the variations from systematic uncertainties on nucleosynthesis and wind ejections (from Limongi & Chieffi 2006)*

These predictions and measurements of abundances of isotopes and elements have been essential to understand both massive stars and their interiors, and how in turn they determine the evolution of galaxy dynamics and composition.

Recent *models of core-collapse supernovae* have shown that the explosion itself occurs relatively late, and is significantly influenced by fallback of stellar-core material and thus processes that reach well beyond the central few 100 km of the star (Fig. 4). Pre-core collapse anisotropies play an important role in the standing-accretion-shock instability (SASI) which is crucial for the success of the explosion. At present, simulations extend the parameter ranges of calculations into those wider arenas, with a corresponding increase in computer resource requirements. Guidance from observations continues to play a key role in advancing our understanding of the important processes that drive cosmic evolution. Spectroscopy of star light will remain essential, also some hope rests on gravitational-wave and neutrino astronomy for more direct probes of the deep interiors of stars. These should be complemented by nuclear-line observations of several key isotopes from supernova/star interiors, to obtain abundances and kinematic information through line shape studies with good energy resolution. For example, the $^{44}$Ti-decay γ-ray lines of Cas A provide important constraints on ejecta from near the mass cut, separating the compact remnant star from ejecta.

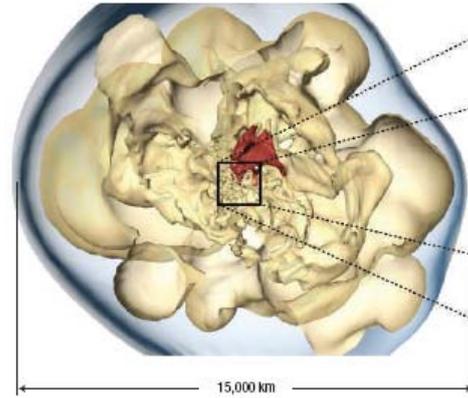

Fig.4: *The processes in the interior of a core-collapse event are complex and subject to 3D effects (from Woosley & Janka 2006)*

**Critical questions concerning our understandings of massive stars:**
- Which nuclear-burning conditions are encountered in late shell burning?
- What is the dynamics of burning shells (flows of fuel and ashes)?
- What is the composition and structure of the star at onset of core collapse?
- Which deviations from symmetry are expected at the onset of core collapse?
- How do the above issues depend on the initial stellar mass and composition?
- How sensitive is explosive nucleosynthesis to rotation and non-sphericity?

Abundances in stellar photospheres are the traditional basis of such studies, complemented by interstellar and intergalactic absorption line studies. Those measurements are confronted with substantial uncertainties of modelling the radiation transport from the stellar core to photosphere; recent revisions of solar metallicity by a factor ~2 illustrate the difficulties of proper interpretations of atomic-line spectra.

**We propose that observations of stellar-surface (photospheric) material (and maybe neutrinos in the future) should be complemented by isotope measurements through γ-ray lines from radioactivity.**

These are intrinsically more direct, originating from nuclear lines aside radioactive decay, which are independent from the thermodynamic state of the gas. Recent experiments have demonstrated that this is feasible. But only the most-intense sources have been seen so far. Sensitivities of γ-ray telescopes should be increased to **generate a sizable sample of nearby stars and young Galactic supernova remnants in the light of $^{26}$Al, $^{60}$Fe, and $^{44}$Ti, as well as a map of Galactic $^{60}$Fe emission.**

Although supernovae are rare events in our Galaxy (~2 per century), measurements of isotopic gamma-ray lines in Galactic supernova remnants, long after the stellar explosion, may provide clues to the physics of the star's interior prior to and during the explosion. This case is exemplified by $^{44}$Ti, already detected in Cas A (a ~360 yr old SN remnant at a distance of 3.4 kpc): the detected line intensity suggests a progenitor star of ~20 $M_\odot$. The kinematics of the radioactive ejecta could be determined by future detectors, and help to constrain the dynamics of the innermost stellar layers during the explosion. Such detectors could also see **$^{44}$Ti in other recent Galactic SN** **events and in SN1987A**, the best studied of all supernovae: Combined with the unprecedented amount of information gathered on that object in all wavelengths, such a detection would provide key (and impossible to obtain otherwise) information on (1) the physical conditions - temperature, density, neutron excess - of the Si-burning shell, during the passage of the shock wave; and (2) the amount of rotation of and fallback onto the nascent compact object. Finally, a systematic survey of the Galaxy in search of $^{44}$Ti in SN remnants would help to answer whether the main producers of $^{44}$Ca ($^{44}$Ti's decay product) in the Galaxy are indeed rare objects - as suggested by the current paucity of such events – or not. The $^{26}$Al radioactivity map of the Galaxy (see below) demonstrates such a measurement of massive-star activity, and already provides a reference to such studies.

**Astrophysical Issue:** **Our Galaxy's Structure and Star Forming Activity**

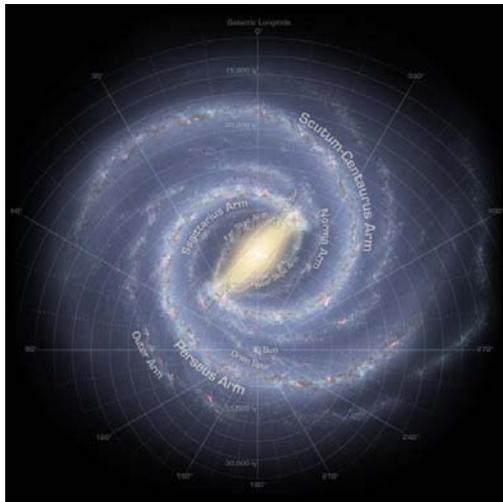 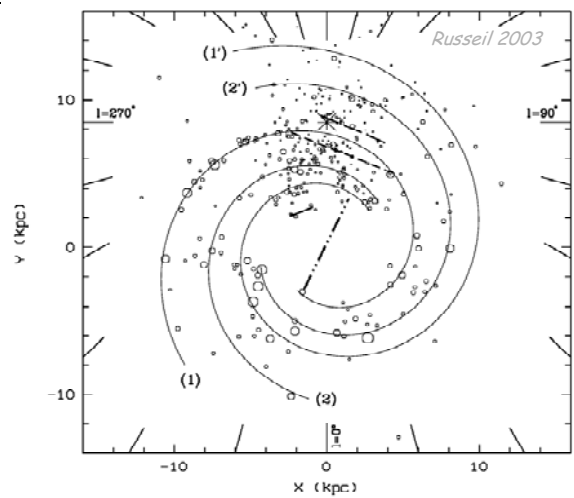

Fig. 5: *An artist's conception of our Galaxy (left, after Spitzer observations), and our current knowledge about the large-scale distributions of star formation (known star-forming complexes, from Russeil 2003). (Both are face-on views, they illustrate the observing difficulties due to our location.)*

Our Galaxy seems the obvious place in the universe where we should know in greatest detail how a spiral galaxy is structured, and how it evolved in time. Formation, evolution, and terminal explosion of stars is the main driver of galaxy evolution. Great efforts are made to understand the conditions under which stars form, and how they eventually end their lives and shed the ashes of nuclear burning into their surroundings. On a grander scale, these questions shape cosmological studies: The most important issue in simulations of galaxy and Large Scale Structure evolution is the treatment of "feedback", and concerns the processes which return mass and energy from stars to the galaxy on time scales of $10^8$ years and shorter. Molecular clouds are disrupted by massive-star activity. Galactic chimneys and fountains have been reported, and could be key targets to understand disk-halo flows and interactions of galaxies.

We propose to utilize penetrating γ-ray surveys to complement studies of these issues from their unique potential to probe hot and tenuous gas near massive stars.

**Our Galaxy** is recognized as a barred spiral galaxy, in which density perturbations develop and propagate through a disk of gas, compress it, and trigger star formation. Towards the central region, a torus of predominantly molecular gas is thought to be the main reservoir of gas from which new stars can form. Observations of nearby galaxies suggest that different types of inner spirals may be formed, as determined by gas densities and rotational characteristics of the disk. Only recently it was recognized that our Galaxy contains a 'bar', a linearly-extended central structure of gas and stars (Benjamin et al. 2005). However, it remains unclear how gas and stars move along the bar itself. Theoretical models (Fig. 6) suggest that in the transition regions between the spiral arms and the bar star formation may be enhanced. At the interface between bar and molecular torus / spiral arms, this should correspond to observable characteristics, which so far have remained undisclosed due to significant occultation. Spiral arms cannot be traced very precisely due to the occulting interstellar ISM. The flow of gas within the bar may be directed inwards towards the center, contributing to feed a central supermassive black hole to gradually grow in mass.

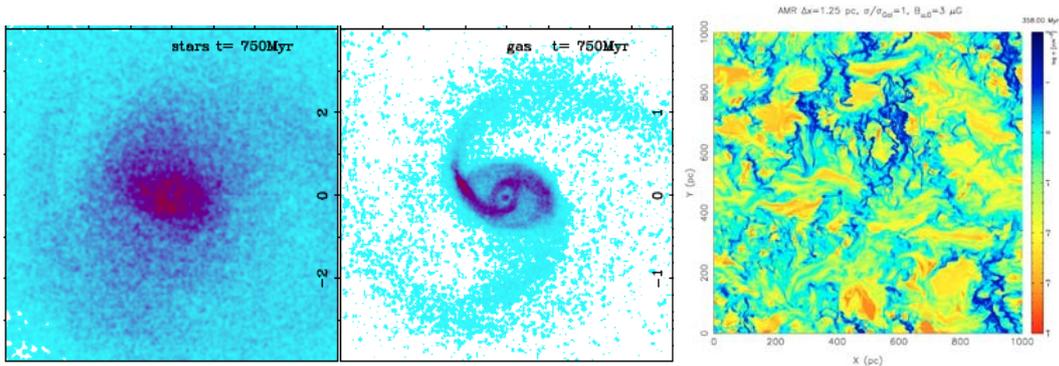

Fig. 6: *Simulations of the evolution of the spatial distributions of stars (left) and gas (center) within a galaxy (from Combes 2008). The bar-like stellar distribution coincides with gas flows in spatially more confined streams. The interstellar medium is found in simulations (right) to be of highly dynamic morphology, massive stars and their supernovae are the agents which lead to transient high-density clouds (yellow/red) and tenuous, hot gas (blue) in between (from de Avillez & Breitschwerdt 2005)*

Rotation profiles vary greatly among galaxies. For our own Galaxy, observations are difficult. CO data suggest a rather large Galactocentric orbital velocity up to 200 km/s within 100 pc of the Galactic Center, but this may be related to peculiar activity of the circumnuclear disk. Observations based on HII regions become sparse and unreliable at galactocentric distances below ~2 kpc. This is the region inside the molecular torus, and in particular the region where bar and inner-spiral flows may deviate substantially from the differential rotation pattern suggested from the large-scale structure determined further out (at radii between 3 and 15 kpc, where rotation settles at a roughly constant value determined by dark-matter distribution). Both from theory and from observations of nearby galaxies it is suggested that inner-galaxy rotation profiles rise steeply with galactocentric radius, or even start from an offset in the 50-100 km/s range.

**Massive stars** are the most-important sources of turbulent energy and newly-formed isotopes ejected into the Interstellar Medium (ISM), they influence their local environments and the Galaxy as a whole. Massive stars occur in groups, evolve together, and their most-massive members eventually destroy the parental molecular cloud with their winds and supernovae. This picture is consistent with a variety of snapshots presented to us in different locations of our- and nearby galaxies, where spatial resolution is sufficient. Important questions remain: How much kinetic energy is available to ISM turbulence? How fast are clouds destroyed? How effective is new-material recycled? Such effects on surrounding ISM can be observed in detail in the nearby star forming regions, such as Orion or Cygnus.

**The ISM.** Massive stars develop strong winds during the 'Wolf Rayet' phase. This, and the ensuing supernova explosion, leads to formation of hot bubbles around these stars (Fig.6), which are filled with the ejected gas, and enriched with newly-formed isotopes. ISM gas will eventually condense and form new interstellar clouds and a new generation of stars. This "feedback" cycle needs to be modeled properly in galaxy-evolution simulations, adding important other astrophysical themes to the already-complex gravitational interactions of gas and stars and dark-matter in interacting/merging galaxies. How massive stars expel and mix their fresh nucleosynthesis products into their surroundings is not fully understood, and difficult to study because the ejected gas is highly ionized and does not radiate efficiently while being hot and tenuous. Present observations mostly rely on the denser boundaries that surround the feedback-driven cavities.

**Critical questions concerning the Galaxy and massive star feedback:**
- What shapes the morphology of (cold and hot) gas & stars in the inner Galaxy?
- How far towards the inner Galaxy can spiral arms be traced?
- How is star formation distributed in the inner Galaxy?
- What is the dynamics of gas as it is driven by star-forming activity?
- Does the outflow from star-forming regions occur through 'chimneys'?
- What are the relevant scales for feedback from massive stars?
- What is the lifetime of molecular clouds after interior star formation?

Cosmological-feedback studies today are characterized by ultra deep surveys of small areas on the sky, of a few deg$^2$ or less, where foreground sources are scarce. While mapping galaxy evolution in (look-back) time is very important, imaging resolution at such large distances limits spatial resolution to the kpc scale, or even coarser. Studies in the nearby universe are needed as a complement. Galaxy Galactic archeology is now undertaken at the necessary precision to unravel the origins of metal-poor, ancient stars in the disk- and halo components of our Galaxy. Here, the elemental abundance signature of the r-Process is the key tool for tracing the activity of *early* star formation and the effects of single, or just a few, supernovae *in the young Galaxy*. Current-day Galaxy studies rely on phenomena related to young-star activity.

**We propose to augment and complement Galactic structure and evolution studies in our Galaxy by tracing massive-star radioactivities.**

Radioactive admixtures are ejected by massive stars into ISM, they decay and emit penetrating γ-rays unaffected by thermo-dynamic conditions of the ISM. This adds the novel window of nuclear γ-ray lines, which are least affected by ubiquitous extinction and probe unique stellar environments and hot, tenuous gas phases that are hardly accessible by other means. **$^{26}$Al in particular has a decay time comparable to the characteristic time of hot-cavity evolution, hence traces the dynamics of hot gas**, therefore such bubble interiors and their evolutions.

A novel determination of inner-Galaxy gas kinematics, based on $^{26}$Al emission from ejecta of young, massive stars, has been advanced with INTEGRAL (Diehl et al. 2006). Penetrating γ-rays as messengers from young-star activity in the inner regions of our Galaxy should be further exploited along those lines, to better understand the interaction between star formation and galactic morphology and its evolution, including constraints on the role of the central supermassive black hole. Deeper imaging with modern γ-ray telescopes can **discriminate star-forming regions and map those also in otherwise difficult regions**. Line spectroscopy using Doppler shifts provides unique **kinematic information on this tenuous, hot gas.** Combining this with kinematic data from cold gas and stars, our present-day Galaxy's morphology can be determined more accurately.

The importance of such mapping of the Milky Way in the light of the long-lived $^{26}$Al line should not be underestimated: it would provide the most accurate picture of recent (last Myr) star formation in the Milky Way, unaffected by biases which severely limit the utility of other tracers. Furthermore, the extension of such an emission vertically to the Galactic plane would put important constraints to models of Galactic chimneys and fountains. Such models are crucial for our understanding of galactic structure and evolution, but observations in other wavelengths cannot address the key issue, namely whether such events indeed transport freshly synthesized metals far away from their sources. The Myr radioactive-decay clock is new information, matching well the time scales of interest for mixing and feedback. γ-rays from $^{60}$Fe as a second, long-lived isotope from probably the same sources provide an opportunity for interesting constraints from γ-ray brightness ratios (Wang et al. 2007).

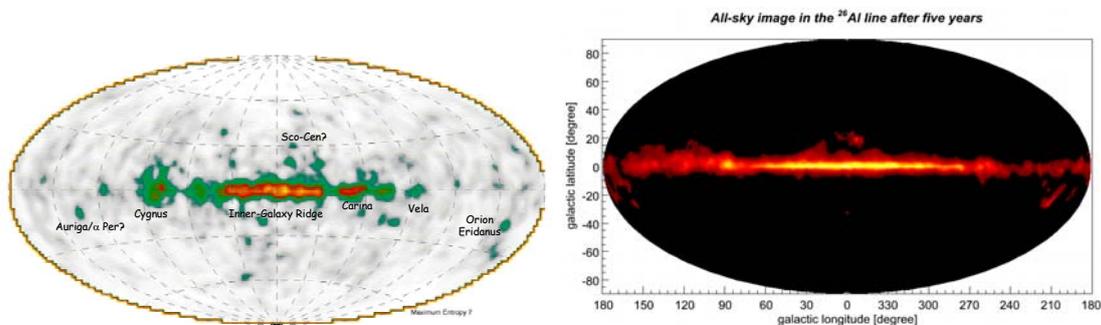

*Fig. 7: The Galaxy's glow in γ-rays from radioactivity has been mapped at the end of the past century in $^{26}$Al (Plüschke et al. 2001). This shows that current massive-star activity is distributed along the plane of the Galaxy with notable irregularities. Deeper measurements of such γ-rays (as anticipated by one such mission proposal, Greiner et al. 2008) can provide new information on massive stars throughout the Galaxy, to be interpreted both in terms of nucleosynthesis and in terms of the Galaxy's morphology.*

A very turbulent state of hot ISM had been suggested from a line-width measurement of the radioactive $^{26}$Al decay gamma-ray line, velocities around 500 km/sec were inferred for the decaying $^{26}$Al nuclei. This would have been hard to reconcile with what we know about the morphology of the ISM and typical sizes of cavities therein (see Fig. 6 right). Improved nuclear $^{26}$Al γ-ray measurements have shown much lower velocities for the hot ISM. These now reach a precision below 100 km/s, an interesting new observational window on the dynamics of turbulent and transient ISM and the issue of stable "phases" of the ISM (Fig 6).

**Astrophysical Issue:         Our Solar-System's Vicinity in our Galaxy**

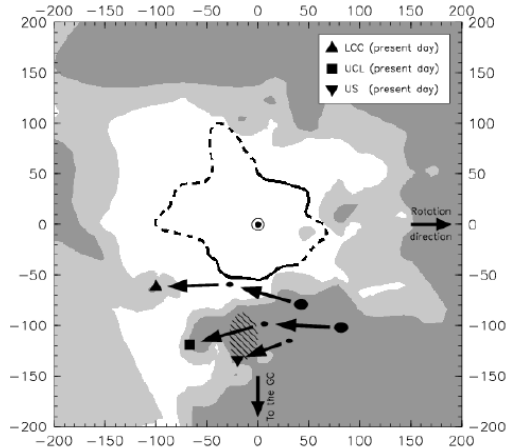

fig. 8: *The solar vicinity is characterized by a hot cavity of gas, with observed streaming motion originating from the general direction of the center of our Galaxy and the nearby Sco-Cen stellar association.*

The Sun is situated within a local interstellar bubble of hot gas, whose origin is largely unknown (Fig. 8). This is an exceptional location, and we should try to understand which events and evolution shaped this local part of the Galaxy. Terrestrial records (Fig. 9) show that an explosion shockwave seems to have swept past the Earth about 3 million years ago. Around that time, also the terrestrial climate underwent a significant change. Climate variations are mediated through cosmic-ray activity driven by nearby supernovae and massive-star activity. Some evidence exists that the nearby stellar association called Scorpius-Centaurus hosts stars whose winds and explosions are responsible for the Sun's environment and its evolution. How special is the Solar interstellar environment? What is the local supernova rate in the Solar vicinity? Do fossil terrestrial records match stellar remnant records?

These questions could be advanced through **imaging measurements of local $^{26}$Al and $^{60}$Fe radioactivity** at sufficient depth and resolution for a mapping wrt. nearby (~100 pc distance or less) sources, i.e., over a large portion of the sky with relatively low surface brightness. The challenge is it: Can we understand, and hence predict, the nearby Myr to Gyrs evolution within our theories and models of how galaxies evolve through massive-star activity? This complements above astrophysical studies, and relates it to our special place, to meteoritic studies, and to the role of radioactivities as heat sources during planet formation.

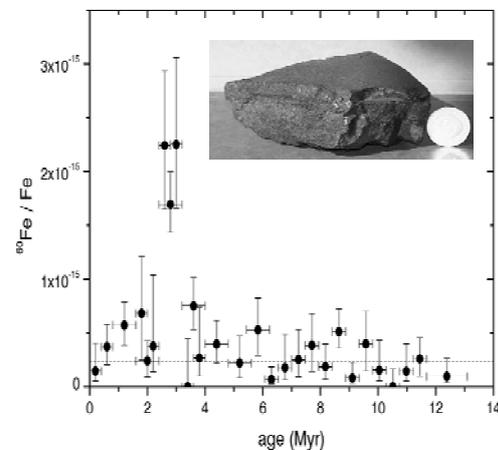

Fig. 9: *Records of massive-star activity in the solar vicinity: The deposit of radioactive $^{60}$Fe in the pacific ocean crust (Knie et al. 2004)*

# Astrophysics with Radioactive Atomic Nuclei

## Proposed Key Activities:

- **Establish a nuclear γ-ray line telescope mission deep enough to obtain a sizeable sample of individual point sources of fresh nucleosynthesis ejecta ($^{26}$Al, $^{44}$Ti, $^{56}$Ni) and positrons (511 keV)**

- **Establish images in nuclear γ-ray lines for extended source regions along the plane of the Galaxy ($^{26}$Al, $^{60}$Fe, 511 keV)**

- **Target key sources with imaging and spectroscopic resolution in nuclear γ-rays to determine the dynamics and spatial characteristics of ejected radioactivities ($^{26}$Al, $^{44}$Ti, $^{56}$Ni)**

- **Implement an imaging γ-ray line survey at modest angular resolution but sufficient depth for tracing individual nearby supernovae back in time for several million years**

## Notes:

Advances in this field would also address several additional topics, such as:

- What is the current core-collapse supernova rate in our Galaxy?
- Are core-collapse supernovae commonly producing $^{44}$Ti in their interiors?
- How much $^{56}$Ni do SNIa produce, and at which velocities?
- How can we exploit the spatial distribution of positron annihilation γ-rays to learn about the sources and propagation of positrons in the Galaxy? ($^{26}$Al and $^{44}$Ti are $e^+$ sources)
- What is the distribution of Galactic novae (from $e^+$ annihilation flashes)?
- Do novae synthesize elements up to Ne (and hence $^{22}$Na)?
- Do AGB stars eject nucleosynthesis ashes such as $^{26}$Al, and how much?
- Where are low-energy (tens of MeV) cosmic-rays produced?
  (from nuclear de-excitation lines emitted from CR/gas interactions)
- Can nuclear de-excitation lines probe high-energy processes in compact accreting binaries?
- What is the kinematics of accelerated particles in solar flares?